\documentstyle[sprocl,psfig]{article}

\def\Journal#1#2#3#4{{#1} {\bf #2}, #3 (#4)}

\def\NPB{{\em Nucl. Phys.} B}
\def\NPA{{\em Nucl. Phys.} A}
\def\PLB{{\em Phys. Lett.}  B}
\def\PRL{\em Phys. Rev. Lett.}
\def\PR{{\em Phys. Rev.}}
\def\PRA{{\em Phys. Rev.} A}
\def\PRC{{\em Phys. Rev.} C}

\def\be{\begin{equation}}
\def\ee{\end{equation}}
\def\bea{\begin{eqnarray}}
\def\eea{\end{eqnarray}}
\def\nc{N_c}

\newcommand{\gas}{g_A^2}

\newcommand{\fps}{F_\pi^2}
\newcommand{\mpi}{m_\pi}
\newcommand{\mpis}{m_\pi^2}
\def\pislash{ {\pi\hskip-0.6em /} }

\def\nopi{ {\rm EFT}(\pislash) }

\def\lamchi{ \Lambda_\chi }
\def\lamqcd{ \Lambda_{\scriptstyle QCD} }
\def\nc{N_c}
\def\infra{{\cal X}}

\def\si{{}^1\kern-.14em S_0}
\def\siii{{}^3\kern-.14em S_1}
\def\piii{{}^3\kern-.14em P_1}
\def\diii{{}^3\kern-.14em D_1}

\begin{document}

\title{NUCLEON-NUCLEON SCATTERING IN \\ THE $1/\nc$ EXPANSION}

\author{SILAS R.~BEANE}

\address{
Department of Physics,
University of Washington,\\
Seattle, WA 98195\\
E-mail: sbeane@phys.washington.edu} 

\maketitle\abstracts{  The nucleon-nucleon $\siii -\diii$ coupled-channel problem is solved
  analytically to leading order in a joint expansion in the quark masses
  and in $1/\nc$. An approximate expression is derived for the $\siii$
  scattering length in the large-$\nc$ limit, and the large-$\nc$ behavior of the deuteron 
  is discussed. }

\section{Introduction}

When the number of colors, $\nc$, of the QCD gauge group is taken large, QCD
simplifies. One may then hope to learn about the real world with $\nc=3$, in an
expansion in $1/\nc$~\cite{'tHooft,Witten}.  The relevance of the large-$\nc$
expansion for nuclear physics was first discussed in the seminal paper by
Witten, where it was argued that the nucleon-nucleon (NN) potential grows
as $\nc$ in the large-$\nc$ limit, and Hartree theory becomes
exact~\cite{Witten}. In a recent revival of interest, it was shown that
large-$\nc$ nuclear interactions are spin-flavor $SU(4)$ symmetric, with Wigner's
phenomenologically successful supermultiplet symmetry following as an
accidental symmetry~\cite{Kaplan1}.  Subsequently, the large-$\nc$ scaling of
the NN potential was analyzed in a general way and large-$\nc$ expectations
were shown to explain generic features of the NN interaction~\cite{Kaplan2}.
Given the levels of complexity that must be unravelled in going from QCD to a
theory of nuclear forces, it is quite remarkable that large-$\nc$
counting rules work so well. Hence large-$\nc$ methods offer an important link
between QCD and nuclear physics. The establishment of such links is, of course,
an important goal for nuclear physics. In this paper, basic
aspects of NN scattering in the large-$\nc$ approximation will be reviewed and
the $\siii -\diii$ NN coupled-channel system will be analyzed in the
large-$\nc$ limit. I will also say some words about the deuteron and the 
bound-state spectrum in the large-$\nc$ limit.

Assuming confinement, in the large-$\nc$ limit the QCD chiral symmetry is spontaneously
broken~\cite{Coleman}. With two light flavors the chiral symmetry breaking
pattern, $U(2)\otimes U(2)\rightarrow U(2)$, gives rise to four Goldstone
bosons, the pions and $\eta$. While the chiral symmetry breaking scale, $\lamchi$,
and the meson masses scale as $\nc^0$, the baryon masses scale as $\nc$.  In
the large-$\nc$ limit the nucleons and $\Delta$s are degenerate and fall into
an infinite-dimensional representation of a contracted $SU(4)$ spin-flavor
symmetry~\cite{Dashen}. In this paper I will assume that $\eta$ and the $\Delta$s
can be integrated out of the low-energy effective field theory (EFT) of
large-$\nc$ QCD in a manner consistent with the large-$\nc$
expansion~\cite{Savage1}. This is, of course, a highly non-trivial assumption.
Consider the resulting EFT involving nucleons and pions below
$\lamchi$ (for reviews, see Ref.~\cite{Be00}). This EFT consists of nucleons
emitting and absorbing pions with an interaction highly constrained by chiral
symmetry. The most general Lagrange density consistent with QCD symmetries is

\begin{eqnarray}
{\cal L}&=& {N^\dagger}\left[ i\partial_t +{\vec\nabla^2}/(2M)\right] N 
-{1\over 2}\left[{C_0^S}(N^\dagger N)^2+{C_0^T}(N^\dagger
  {\vec\sigma}N)^2\right]\nonumber\\
&+&{\fps\over 4}\, Tr\, \partial_\mu \Sigma^\dagger \partial^\mu \Sigma +
\omega{\fps\over 2}\, Tr\, {\cal M}_q \left(\Sigma^\dagger +\Sigma\right) +
g_A N^\dagger {\vec A}\cdot{\vec\sigma}N+\ldots
\label{lagrangian}
\end{eqnarray}
where ${\vec A}$ is the axial-vector current and the ellipses represent
operators with more derivatives, more powers of the quark mass matrix,
and more nucleon fields. In the large-$\nc$
limit, $F_\pi\propto\sqrt{\nc}$, $g_A\propto{\nc}$ and
$\omega\propto{\nc^0}$~\cite{Witten}. The four-nucleon operators scale as
${C_0^T}\propto{1/\nc}$ and ${C_0^S}\propto\nc$~\cite{Kaplan1}.  When
considering the interactions of more than one nucleon, the ordering of the
operators in Eq.~(\ref{lagrangian}) by way of a power-counting scheme is
nontrivial. This is especially so in the $\siii-\diii$ coupled-channel of NN
scattering in an EFT with pions, which I will focus on in this paper. The presence
of the deuteron, a near-threshold bound state, and the highly-singular tensor
force imply complicated nonperturbative physics.

\section{NN Scattering at Large-$\nc$}

Defining a sensible large-$\nc$ limit for NN scattering is nontrivial because
the nucleon mass grows as $\nc$. There are several limits we may consider. 
The momentum-space Schr\"odinger equation is 

\begin{equation}
 \left({{\hat{p}^2}\over{M}} + \hat{V}\right)\, |\Psi>=E\, |\Psi >
\ ,
\label{eq:Hamil}
\end{equation}
where $M$ is the nucleon mass.
At fixed velocity, the momentum transfer scales as $\nc$. Since the potential
also scales as $\nc$, an overall factor of $\nc$ scales out of the Hamiltonian
and Hartree theory becomes exact~\cite{Witten}. Evidently there is no EFT description in
this limit as characteristic momenta quickly overcome $\lamchi$ and a
description in terms of quarks and gluons becomes appropriate.  At fixed
momentum transfer, which is the limit relevant to EFT\footnote{In principle,
the momentum transfer can scale as an inverse power of $\nc$ in the EFT, but
then one is constrained to threshold.}, the nucleon kinetic
energy is suppressed by $1/\nc$ while the potential energy grows as $\nc$. In
this limit, the nucleons are always below threshold for
scattering~\cite{Coleman2}. The static nucleons settle at the bottom of the
potential well and nuclear matter forms a classical crystal~\cite{Kaplan2}.
The absence of a suitable large-$\nc$ limit to describe scattering at fixed
momentum transfer poses something of a dilemma. One way out is to focus on
the large-$\nc$ scaling properties of the NN potential and then to compare
the resulting predictions with modern phenomenological NN potentials that
fit phase shifts to high precision~\cite{Kaplan1,Kaplan2}. This has proved
highly successful. The question of whether the fixed momentum transfer limit 
exists has been called into question in interesting recent
work~\cite{Banerjee,BelCo} which points out that there would appear to
be violations of the counting rules for the potential arising from diagrams with
multiple meson exchanges. This is an important unsolved puzzle.

\section{The (Un)Coupled Channel Problem at Large-$\nc$}

\noindent The most economical power-counting scheme for the $\siii -\diii$ coupled
channels is an expansion about the chiral limit~\cite{Be01}. 
The pion-exchange part of the NN force in the chiral limit is purely tensor,

\begin{equation}
V_T(r)\ =\ -{{3\alpha_\pi}\over{r^3}}\ + \ {\cal O}(\mpis ) \ ,
\label{eq:tensorchirallimit}
\end{equation}
where

\begin{equation}
{\alpha_\pi}={{\gas}\over{16\pi\fps}}.
\label{eq:alphadefined}
\end{equation}
At leading order in the EFT expansion there is a local four-nucleon operator
containing the effect of non-pionic short-distance physics which renormalizes
the tensor force at short distances. The coefficient of this operator is
${C_0}\equiv{C_0^S}+{C_0^T}={C_0^S}+{\cal O}(1/\nc)$~\cite{Kaplan1}. I choose
to regulate this coordinate-space delta-function using a square well,

\begin{equation} 
{C_0}\;\delta^{(3)} (r)\rightarrow {{3{C_0}\; \theta (R-r)}\over{4\pi
    R^3}}\equiv {V_0}\;\theta (R-r) \  ,
\label{eq:ctovsinglet}
\end{equation} 
with strength $V_0$ and width $R$.
The $S$-wave and $D$-wave components of the $J=1$ system, $u(r)$ and $w(r)$
respectively, are coupled by the tensor potential, $V_T$, defined in
Eq.~(\ref{eq:tensorchirallimit}).  The long-distance potential outside the square well is

\begin{equation} {\cal V}_L(r)= \left(
\begin{array}{cc}   
  0  & -2\sqrt{2}\; M V_T(r)\\               
-2\sqrt{2}\; M V_T(r) & 2M V_T(r)-{6/{r^2}}
          \end{array}
\right),
\end{equation}
while the short-distance potential inside the square well is

\begin{equation}
{\cal V}_S(r)=
\left(
\begin{array}{cc}   
-M V_0  & 0\\               
0     & -M V_0-{6/{r^2}}
          \end{array}
        \right) \ \ .  
\end{equation} 
Defining $\Psi$ to be

\begin{equation}
\Psi (r)= \left(\begin{array}{c}
            u(r)\\
            w(r)
            \end{array}\right)
          \ \ \ , 
\end{equation} 
the regulated Schr\"odinger equation takes the compact form

\begin{equation}
          {\Psi'' (r)}+ \left( k^2 + {\cal V}_L(r)\theta (r-R) +{\cal
              V}_S(r)\theta (R-r)\right) \Psi=0.  
\label{eq:schro}
\end{equation} 
One may ask whether there is a well-defined large-$\nc$ limit.  In
Eq.~(\ref{eq:schro}), $M V_T$ and $M V_0$ scale as ${\nc^2}$, while the angular-momentum
barrier scales as ${\nc^0}$. In the EFT we treat the momentum transfer as
fixed ($\nc$-independent). The total
center-of-mass-energy is then suppressed by $1/{\nc^2}$ compared to the interaction
energy. Here we will focus on this rather peculiar large-$\nc$ limit.  At
leading order in the $1/{\nc}$ expansion, the
coupled channel problem can be diagonalized via the similarity transformation
$\Psi \rightarrow {S^{-1}}{\Psi}$ where

\begin{equation}
S=\left(
\begin{array}{cc}   
\sqrt{2}\  & \ -1/\sqrt{2}\\               
1\     & \ 1
          \end{array}
        \right) \ \ .  
\end{equation} 
Clearly $\siii-\diii$ mixing is suppressed in the large-$\nc$ limit. 
The general solution of Eq.~(\ref{eq:schro}) in the large-$\nc$ limit 
is a linear combination of Bessel functions. Keeping the leading contributions I find

\begin{eqnarray}
{u}(r)&=&\sqrt{2}\ A\ r^{3/4}\cos\left( 2\sqrt{{{6 \alpha_\pi M}\over r}}+\phi_0 \right)
-{B\over\sqrt{2}}\ r^{3/4}\cosh\left( 2\sqrt{{{12 \alpha_\pi M}\over
      r}}+\delta_0 \right)
\nonumber\\
{w}(r)&=& A\ r^{3/4}\cos\left( 2\sqrt{{{6 \alpha_\pi M}\over r}}+\phi_0 \right)
+ B\ r^{3/4}\cosh\left( 2\sqrt{{{12 \alpha_\pi M}\over
      r}}+\delta_0 \right),
\label{decoupled}
\end{eqnarray}
where $\phi_0$, $\delta_0$, $A$ and $B$ are determined from boundary
conditions. Note that the repulsive hyperbolic solutions grow exponentially
with $\nc$. Since these solutions cannot match to a power law in $\nc$, we find
$B=0$ in the large-$\nc$ limit. We then have

\begin{equation} 
{u}(r)=\sqrt{2}\ A\ r^{3/4}\cos\left( 2\sqrt{{{6 \alpha_\pi M}\over r}}+\phi_0 \right)
=\sqrt{2}\ {w}(r),
\label{decoupled2}
\end{equation}
or

\begin{equation} 
{{{u}(r)}\over{{w}(r)}}=\sqrt{2}+O\left({1\over{\nc^2}}\right)
\label{decoupled3}
\end{equation}
and the $S$-wave and $D$-wave wavefunctions oscillate {\it in phase}, with a
ratio of $\sqrt{2}$. If there are nodes in the wavefunctions, then they must overlap.
It is remarkable that such detailed information about
nuclear wavefunctions can be deduced directly from QCD. Notice that the $1/\nc$
expansion is equivalent to the short-distance expansion. 
Eq.~(\ref{decoupled3}) is a remarkably accurate representation of the
short-distance part of deuteron wavefunctions in conventional NN models with both
hard- and soft-cores whose long-range part is governed by pion
exchange~\cite{Sprung}. At large distances in the real world, the angular
momentum barrier is not negligible. With $M{\alpha_\pi}\equiv
1/{\Lambda_{NN}}$, the tensor force dominates the angular momentum barrier for 
distances $r\ll 1/{\Lambda_{NN}}$, which are outside the validity of the EFT
description. It might prove useful to assign a more realistic large-$\nc$
scaling to the angular momentum barrier and treat it as a perturbation using 
the WKB approximation, which becomes arbitrarily accurate in the large-$\nc$ limit.

The phase $\phi_0$ is determined by matching logarithmic derivatives
of the interior and exterior wavefunctions,

\begin{equation} 
{\sqrt{-M V_0}R}\cot{\left({\sqrt{-M V_0}R}\right)}={3\over 4} + 
\sqrt{{6M{\alpha_\pi}}\over R}
\tan \left(2\sqrt{{6M{\alpha_\pi}}\over R}+\phi_0\right).
\label{eq:logderivativeV0}
\end{equation}
This equation determines the renormalization-group flow of the four-nucleon
contact operator~\cite{kiddies,Be01}. The multi-branch structure of the right side of
Eq.~(\ref{eq:logderivativeV0}) is a consequence of the nonperturbative treatment
of the tensor force. Treating
$R\equiv 1/{\Lambda}$ as a Wilsonian cutoff, we see that the tensor force
becomes perturbative when ${\Lambda /{\Lambda_{NN}}}\ll 1$.  In the large-$\nc$
limit, ${\Lambda /{\Lambda_{NN}}}\propto{\nc^2}\gg 1$ and the tensor force
becomes nonperturbative at arbitrarily small momentum scales. This suggests an
interesting bound-state spectrum in the large-$\nc$ limit.

\section{A Deuteron in the Large-$\nc$ Limit?}

It is an intriguing mystery that QCD has a single scale, $\lamqcd$, which is
${\cal O}(100~{\rm MeV})$, while characteristic nuclear binding energies are
${\cal O}(1~{\rm MeV})$. In particular, the deuteron binding energy, $-B$, is
$B=2.224575~{\rm MeV}$. The problem might not actually be this severe. The
tensor force in the chiral limit is governed by a single scale, $F_\pi$. If we neglect geometrical
factors and assume that the momenta of the bound nucleons is $\sim F_\pi$, then
we would expect a binding energy of order $F_\pi^2/M\sim 10~{\rm MeV}$. Of
course this argument breaks down as we vary $\nc$, since the axial coupling
grows with $\nc$. Before considering the bound state spectrum in the pionful
EFT, we will consider a simplified scenario.

At distance scales much larger than the pion Compton wavelength, the pion can
be integrated out and the EFT simplifies considerably~\cite{We90}.  In $\nopi$ observables
are determined purely by four-nucleon contact operators~\cite{Be00}.  In the $\siii-\diii$
coupled-channel there is a single momentum-independent interaction, $C_0\propto\nc$. We
will assume that this operator is dominant at low energies in the large-$\nc$
limit. The $\siii$ scattering length is then

\begin{equation}
a=R\left(1-{{\tan{\left({\sqrt{-M V_0}R}\right)}}\over{{\sqrt{-M V_0}R}}}\right)
\longrightarrow{{M C_0}\over{4\pi}},
\label{eq:scatt}
\end{equation}
where we use a square-well regulator and the arrow indicates that
renormalization has been performed and $C_0$ is a renormalized quantity.
Hence $a\propto{\nc^2}$.  If the sign of $C_0$ is positive, then there is a
bound state with binding energy

\begin{equation}
B= {{16\pi^2}\over{{M^3}{ C^2_0}}}.
\label{eq:scatt2}
\end{equation}
Hence, $B\propto{1/\nc^{5}}$. What do we expect if we include pions in the EFT?
We have seen that the tensor force dominates the potential at long distances.
Hence we can write the Schr\"odinger equation in coordinate space as 

\begin{equation}
\left({{{\nabla}^2}\over{\nc\bar{M}}} 
-{{3{\nc}{\bar{\alpha}_\pi}}\over{r^3}}\right)\, |\Psi>=E\, |\Psi > \ ,
\label{eq:Hamil2}
\end{equation}
where the $\nc$ dependence has been scaled out and the barred quantities are $\nc$-independent.
We can scale out the $\nc$ dependence through a rescaling
of the coordinate, $r\rightarrow{\nc^2}{\bar{r}}$. The eigenvalue equation then
becomes

\begin{equation}
{1\over{\nc^5}}\left({{{\bar{\nabla}}^2}\over{\bar{M}}} 
-{{3{\bar{\alpha}_\pi}}\over{\bar{r}^3}}\right)\, |\Psi_B>\ =
{B}\, |\Psi_B>\ ,
\label{eq:Hamilbound}
\end{equation}
and we again find $B\propto{1/\nc^{5}}$. This argument can be formulated in a
more physical way as follows~\cite{markus}. We can write the potential as

\begin{equation}
V_T\sim {{\nc}\over{{r_B}^3}},
\label{eq:tensorchirallimitlargen}
\end{equation}
where $r_B$ can be loosely interpreted as the size of the bound state and I have
assumed that all dimensional parameters are given by powers of $\lamqcd$,
which is defined to be unity.  Now assume that there is a bound state in the
$\siii$ channel at large-$\nc$.  The presence of a bound state requires a
balance between the kinetic energy, $\nc {v_B^2}/2$, and the potential energy,
and therefore we expect $V_T \sim \nc {v_B^2}$. Here $v_B$ is the velocity of
the bound state. As required for consistency, $\nc$ drops out of this
relation and we arrive at ${v_B}^2\sim {r_B}^{-3}$. This is simply a
statement of the virial theorem. Since the nucleon momentum is $\nc {v_B}$, it
follows from the uncertainty principle that ${v_B}\sim (\nc {r_B})^{-1}$. Again
using the virial theorem, we find

\begin{equation}
{r_B}
\propto {\nc ^2} \qquad\qquad {v_B}\propto 1/\nc ^{3}  \qquad\qquad  B \propto
1/{\nc^{5}}\ ,
\label{eq:dsizevel}
\end{equation}
and we have a bound state that is large, slow, and shallow in the large-$\nc$
limit. Hence our naive scaling expectations in the pionful EFT agree with those from
$\nopi$. Unfortunately both of these arguments would appear to be too naive as
they ultimately rely on the virial theorem constraint that the coordinate scales
as ${\nc^2}$, or, equivalently, that the momentum transfer scales as
$1/{\nc^2}$. As we will see below, a more detailed study in the pionful theory
suggests that there are an infinite number of bound states in the large-$\nc$ limit,
with the ground state binding energy increasing as a complicated function of $\nc$.
This last point is not surprising, given the large-$\nc$ wavefunctions found in the previous
section. If one acts with the kinetic energy operator on the wavefunction of
Eq.~(\ref{decoupled2}), one finds a complicated (harmonic) dependence on $\nc$.

\section{The Bound-State Spectrum}

\begin{figure}[t]
\hspace{1.3cm}
\psfig{figure=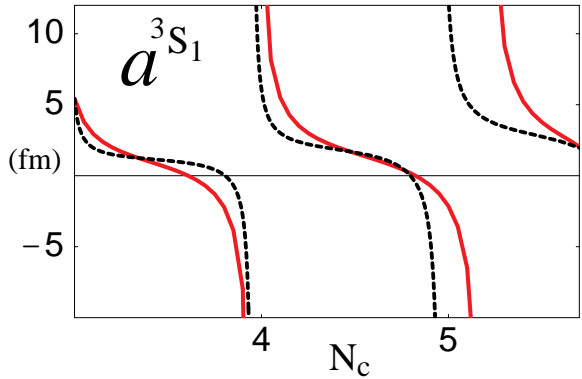,height=2in}
\caption{
The $\siii$ scattering length as a function of $\nc$. 
The solid line is a numerical solution of the Schr\"odinger equation
with a small but finite pion mass and with $R=0.375~{\rm fm}$ and
a square-well depth of $M V_0 =-2.07\times 10^6 ~{\rm MeV}^2$ tuned
to the physical scattering length at $\nc =3$ ($\phi_0= -7.6$). The dashed line
is the approximate analytic solution, Eq.~(\ref{eq:scattapp}) with
$\phi_0= -7.6$ and $\infra =0.97~{\rm
  fm}$ which reproduce the physical scattering length at $\nc =3$.}
\label{fig:scattpanel}
\end{figure}

\begin{figure}[t]
\hspace{1.3cm}
\psfig{figure=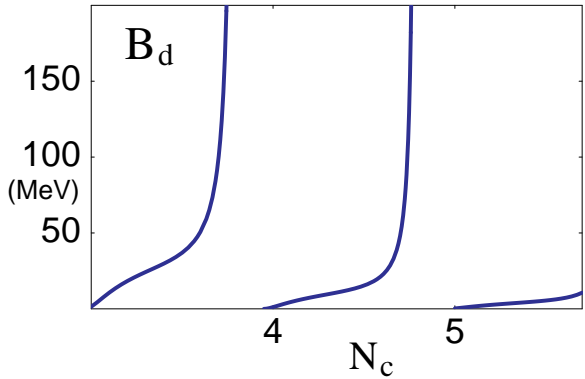,height=2in}
\caption{
The deuteron binding energy as a function of $\nc$ abstracted from the
approximate analytical solution, Eq.~(\ref{eq:scattapp}), and verified
numerically along the plateaus.}
\label{fig:bdpanel}
\end{figure}

One way of studying the bound-state spectrum is via the scattering length.
Although a scattering length is strictly speaking not defined for a $1/r^3$ potential~\cite{kiddies}, we can
impose an infrared cutoff, $\infra$, on $V_T$ and match to a wavefunction at
$r>\infra$ of the form $r-{a^{\siii}}$. This gives

\begin{equation} 
{a^{\siii}}=
{{\infra\left( 
-1+4\sqrt{{6M{\alpha_\pi}}/\infra}\tan \left(2\sqrt{{6M{\alpha_\pi}}/\infra}+\phi_0\right)
\right)}\over{
\left(
3+4\sqrt{{6M{\alpha_\pi}}/\infra}\tan \left(2\sqrt{{6M{\alpha_\pi}}/\infra}+\phi_0\right)
\right)}}.
\label{eq:scattapp}
\end{equation}
The purpose of this formula is to give analytical understanding of the bound state
spectrum in the large-$\nc$ limit. We would expect the infrared cutoff to take
a value around $1/\mpi$. Choosing an ultraviolet cutoff $R=0.375~{\rm fm}$ and
a square-well depth of $M V_0 =-2.07\times 10^6 ~{\rm MeV}^2$ which reproduce
the scattering length for the physical value of $\nc =3$ in the full problem,
gives $\phi_0= -7.6$. I then choose the infrared cutoff 
$\infra =0.97~{\rm   fm}$ to reproduce the scattering length in the 
approximate formula. In order to explore the bound state spectrum as a function
of $\nc$, I naively scale all quantities in the approximate formula and in the
exact numerical result by the expected large-$\nc$ scaling.
The approximate formula for the scattering length and the
exact numerical result are plotted as a function of $\nc$ in Fig.~(\ref{fig:scattpanel}).
The approximate formula is reasonably accurate in
the vicinity of $\nc=3$ (for several branches of the tangent). 
The corresponding plot of the binding energy is given
in Fig.~(\ref{fig:bdpanel}). The plotted expression is derived from 
the approximate formula for the scattering length using effective
range theory, and has been verified numerically along the plateaus using the exact solution.
With the given choice of cutoff and square-well depth,
there are no nodes in the wavefunction and the deuteron (at $\nc =3$) is, by
construction, the ground state.
As $\nc$ is increased from its physical value, the
deuteron binding energy grows as a complicated function of $\nc$ and quickly
passes beyond $\lamchi$ and out of the EFT. 
There is then no bound state in the EFT. As $\nc$ approaches 4, a new bound
state, the first excited state,
appears at threshold and then undergoes the same decoupling as $\nc$ is
increased. This cycle repeats itself an infinite number of times in the
large-$\nc$ limit. The excited levels can be studied in more detail through the
singularities of the scattering length, using the approximate formula,
Eq.~(\ref{eq:scattapp}). One can estimate the density of eigenstates~\cite{PP} 
for the $1/r^3$ potential using WKB; with excitation number, $n$, one finds
$|E_n|\sim n^6 \lamqcd$ with $n=1,2\ldots$. Therefore as
$\nc\rightarrow\infty$ there are an infinite number of bound states,
infinitely dispersed in energy. But there would appear never to be more
than one bound state present in the EFT.

\section{Caveats}

There are many caveats to the large-$\nc$ picture of NN scattering presented in
this paper. Among them is the absence of the $\Delta$ degrees of freedom.  NN
scattering becomes a significantly more complicated coupled channel problem
when the $\Delta$s are degenerate with the nucleons~\cite{Savage1}. It would be
interesting to repeat the analysis presented here with the complete contracted
spin-flavor $SU(4)$ multiplet.  A more serious problem lies in the fact that
the large-$\nc$ limit in which the momentum transfer is held fixed does not
experience scattering, and potentially does not exist as a sensible limit of
QCD~\cite{Banerjee,BelCo}. One may also seriously question the procedure of
varying $\nc$ slightly away from its physical value to deduce information about
the large-$\nc$ limit. It is conceivable that the true large-$\nc$
Schr\"odinger equation behaves quite differently.  On a related note, there are
several issues of noncommutativity of limits.  First there is the issue of
whether to take the large-$\nc$ limit at the level of the Schr\"odinger
equation, or at the level of its solution.  There is also the issue of whether
the chiral limit commutes with the large-$\nc$ limit; note that
the tensor force in the chiral limit is highly singular at all distances, and
this has an important effect on the bound state spectrum.
Anthropic arguments suggest that the fine-tuned nature of the deuteron binding
energy is essential to the emergence of life. The results
presented in this paper suggest that the large-$\nc$ limit can be relevant to
nature only at discrete values of $\nc$ where there continues to be a bound
state with deuteron quantum numbers and binding energy within anthropic bounds.
Whether this be the ground state or the $\nc$th excited state is irrelevant in
the EFT. Other (large) values of $\nc$ --away from the plateaus in
Fig.~(\ref{fig:bdpanel})-- would in all likelihood not support the intelligent
life necessary to think about large-$\nc$ QCD.

\section*{Acknowledgments}
I am grateful for valuable conversations with David Kaplan,
Markus Luty, Martin Savage and Bira van Kolck. Special thanks to the
organizers, particularly Rich Lebed, for a stimulating workshop. 
This research was supported in part by the DOE grant DE-FG03-97ER41014.

\end{document}